# Domain structure and perpendicular magnetic anisotropy in CoFe/Pd multilayers using off-axis electron holography*

Desai Zhang[a)]

School of Engineering for Matter, Transport, and Energy, Arizona State University, Tempe, AZ 85287
Email: dzhang28@asu.edu

Justin M. Shaw

Electromagnetics Division, National Institute of Standards and Technology, Boulder, CO 80305
Email: justin.shaw@nist.gov

David J. Smith

Department of Physics, Arizona State University, Tempe, AZ, 85287
Email: DAVID.SMITH@asu.edu

Martha R. McCartney

Department of Physics, Arizona State University, Tempe, AZ, 85287
Email: Molly.McCartney@asu.edu

______________________________

[a)] Corresponding author, phone: 001 480 4526664

**Abstract:**

Multilayers of $Co_{90}Fe_{10}$/Pd with different bilayer thicknesses, have been deposited by dc-magnetron sputtering on thermally oxidized Si wafers. Transmission electron microscopy showed that the highly textured crystalline films had columnar structure, while scanning transmission electron microscopy and atomic force microscopy respectively indicated some layer waviness and surface roughness. The magnetic domain structure and perpendicular magnetic anisotropy (PMA) of the $Co_{90}Fe_{10}$/Pd multilayers were investigated by off-axis electron holography and magnetic force microscopy. The $Co_{90}Fe_{10}$ layer thickness was the primary factor determining the magnetic domain size and the perpendicular magnetization: both decreased as the thickness increased. The strongest PMA was observed in the sample with the thinnest magnetic layer of 0.45 nm.

**Highlight:**

- Multilayers of $Co_{90}Fe_{10}$/Pd with different bilayer thicknesses were deposited by dc-magnetron sputtering on thermally oxidized Si wafers.
- Off-axis electron holography investigation confirmed that the $Co_{90}Fe_{10}$ layer thickness was the primary factor determining the magnetic domain size and the perpendicular magnetic isotropy (PMA): both decreased as the thickness increased.
- Other features revealed by (scanning) transmission electron microscopy had little impact on magnetic properties.

# 1. Introduction

Magnetic thin films and nanostructures that are magnetized perpendicular to their surface are essential to many developing technologies, including spintronics devices [1] and patterned media [2-3], especially because of the need to maintain thermal stability as device dimensions are reduced deeper into the nanoscale. Multilayers or superlattice structures consisting of alternating ferromagnetic and nonmagnetic layers are highly suitable for these applications due to their tunable perpendicular magnetic anisotropy (PMA) and saturation magnetization [4-6]. The magnetic microstructure is crucial for developing potential devices, and such multilayered structures have been extensively studied using different characterization techniques [7-13]. For example, magnetic force microscopy (MFM) applied to samples consisting of Co/Pd multilayers showed striped domain structure [9-10], and X-ray magnetic circular dichroism (XMCD) revealed changes in the domain structure in different layers of NiFe/Au/Co/Au multilayers [11]. Magnetic optical Kerr (MOKE) microscopy has also been used to reveal magnetic domain formation and switching behavior [12], and it was reported that PMA was present in amorphous CoSiB/Pt multilayers that had similar striped domains [13]. In this present work, off-axis electron holography has been used to investigate the PMA and magnetic domain structure of $Co_{90}Fe_{10}$/Pd multilayers of different layer thicknesses. Transmission electron microscopy (TEM), and high-angle annular-dark-field (HAADF) scanning transmission electron microscopy (STEM) were used to determine the microstructure, and atomic force microscopy (AFM) together with magnetic force microscopy (MFM) were used to compare the surface morphology and domain structure of the films.

# 2. Experimental details

The multilayer system consisted of $Co_{90}Fe_{10}$/Pd multilayers with the basic structure of Ta/Pd/[$Co_{90}Fe_{10}$/Pd] ×12/Pd, where the thicknesses of the magnetic layers were $t_{CoFe}$ = 0.45 nm, 0.8 nm or 1 nm, and the ratio of thicknesses between the $Co_{90}Fe_{10}$ and Pd layers was kept fixed at 1 : 2. All samples were prepared at room temperature by dc magnetron sputter deposition at an Ar pressure of ~ 66 mPa (0.5 mTorr) onto oxidized Si(001) substrates. Deposition rates were calibrated using x-ray reflectivity (XRR), which indicated a maximum drift in deposition rates of 3% throughout the study. All samples contained 3-nm Ta / 3-nm Pd seed layers, and a final 3-nm Pd capping layer was also used to prevent oxidation.

Samples suitable for cross-sectional (S)TEM observations were prepared by cleaving and mechanical wedge-polishing, followed by 2.5-keV Ar-ion-milling at liquid-nitrogen temperature to create electron-transparent areas. TEM and STEM images were recorded using a JEOL JEM-2010F[†] equipped with a thermal-field-emission electron gun operated at 200 keV. Off-axis electron holography studies were performed using a Philips-FEI CM200 FEG TEM[†] operated at 200 keV and an FEI Titan 80-300 ETEM[†] operated at 300 keV. The latter two microscopes were each equipped with an electrostatic biprism and a small (Lorentz) mini-lens located below the lower pole piece of the objective lens, enabling the samples to be studied in field-free conditions. A positive biprism voltage of ~100 V, corresponding to an interference-fringe spacing of ~2 nm, was typically used to record electron holograms in the Lorentz imaging mode using a CCD camera. In the field-free imaging condition, the presence of PMA was revealed by magnetization perpendicular to the plane of the multilayer interfaces. To confirm the magnetic easy axis of the multilayers, all TEM samples were also magnetized *in situ* using magnetic fields of close to 2 T, with the direction of the applied field parallel to the multilayer interfaces (i.e. in the film plane). This external magnetic field was created using the objective lens of the microscope and was removed before the holograms were recorded.

# 3. Results and discussion

## 3.1 Surface morphology

AFM was used to check the sample surfaces after deposition, and results for the three samples are shown in the left column of Fig. 1. Granular surface features were visible for all samples with the sizes of these features increasing for larger $t_{CoFe}$. Statistical analysis of the images indicated that the average grain sizes of the samples were 23 nm, 26 nm and 29 nm for $t_{CoFe}$ = 0.45 nm, 0.8 nm and 1 nm, respectively. MFM scans to reveal out-of-plane magnetic domain structure were also made after applying in-plane magnetic fields of 1.1 T, and the corresponding results for each sample are shown at the right side of each AFM image. By comparing images for $t_{CoFe}$ = 0.45nm and 0.8nm, it is clear that increasing $t_{CoFe}$ changes the magnetic domain structure and decreases the domain size. A striped domain structure in Fig. 1(e) is apparent for $t_{CoFe}$ = 0.8nm but no out-of-plane domain structure is visible in Fig. 1(f), corresponding to the sample with $t_{CoFe}$ = 1 nm.

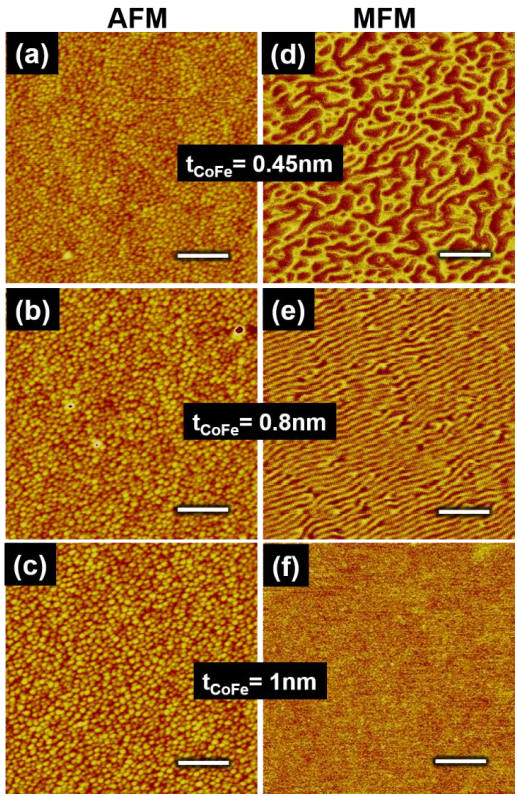

**Fig. 1.** AFM images (left), and corresponding MFM images (right), from the CoFe/Pd multilayer surfaces. Scale bar for AFM images = 1 μm, for MFM images = 200 nm.

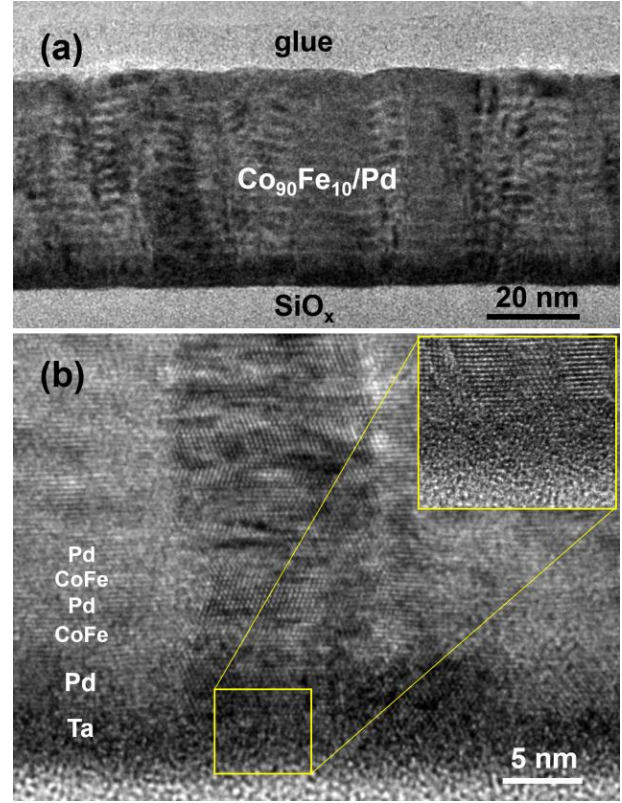

**Fig. 2.** (a) Cross-section TEM image of CoFe/Pd multilayer sample with $t_{CoFe}$ = 0.8 nm showing columnar morphology; (b) HRTEM image showing bottom part of multilayer for $t_{CoFe}$ = 0.8 nm showing textured crystal grains with random in-plane orientations. Enlargement shows layer just above $SiO_x$.

3.2 Crystalline structure

Figure 2(a) is a bright-field TEM image showing a cross section of the $Co_{90}Fe_{10}$/Pd multilayers for the sample with $t_{CoFe}$ = 0.8 nm. In this diffraction-contrast image, the separate $Co_{90}Fe_{10}$ alloy and Pd layers are visible as roughly horizontal regions of lighter and darker contrast, respectively. However, the contrast within each layer is not uniform, indicating that the multilayers are not well defined and/or have considerable thickness variation and polycrystallinity. Moreover, the image contrast in the direction perpendicular to the substrate surface indicates that the film has a columnar morphology, which causes some surface roughness. As indicated in Fig. 2(a), these columns and surface bumps have dimensions that roughly correspond to the average width of the granular features visible in AFM scans such as Fig. 1(b). High-resolution phase-contrast images, such as Fig. 2(b), indicate that the columns are crystalline with pronounced in-plane texture. A narrow region with dark contrast, corresponding to the Ta and Pd seed layers, is visible between the $SiO_x$ and metal multilayers. This layer can be observed more clearly in the enlargement inserted in Fig. 2(b), and shows some slight signs of crystallinity. The samples with larger $t_{CoFe}$ had reduced layer waviness and surface roughness relative to the sample with $t_{CoFe}$ = 0.45 nm.

The waviness of the layers is more apparent using STEM HAADF imaging since the image contrast in this mode is highly dependent on atomic number. Figures 3(a) and (b) compare HAADF images from the samples with $t_{CoFe}$ = 0.45 nm and 1 nm, respectively. The multilayer structure is clearly visible, and the thicknesses of the layers are consistent with the nominal values. The multilayers become increasingly wavy away from the substrate and the interfaces seem to be less well defined, while the sample with $t_{CoFe}$ = 0.45 nm shows greater relative waviness than the sample with $t_{CoFe}$ = 1 nm. Line profiles of image intensity are shown in Figs. 3(c) and (d). Following the deconvolution method for determining compositional variations used by Crozier and colleagues [14-15], intensity simulations were carried out, and these are shown as inserted bold lines in Figs. 3(c) and (d). In these simulations, the nominal elemental compositions and profiles of the $Co_{90}Fe_{10}$/Pd superlattices were used. The differences between the experimental and simulated profiles indicate that some interdiffusion has occurred at the interfaces. Since all samples were grown under the same conditions, this interdiffusion will have a relatively greater effect on samples with smaller layer thicknesses.

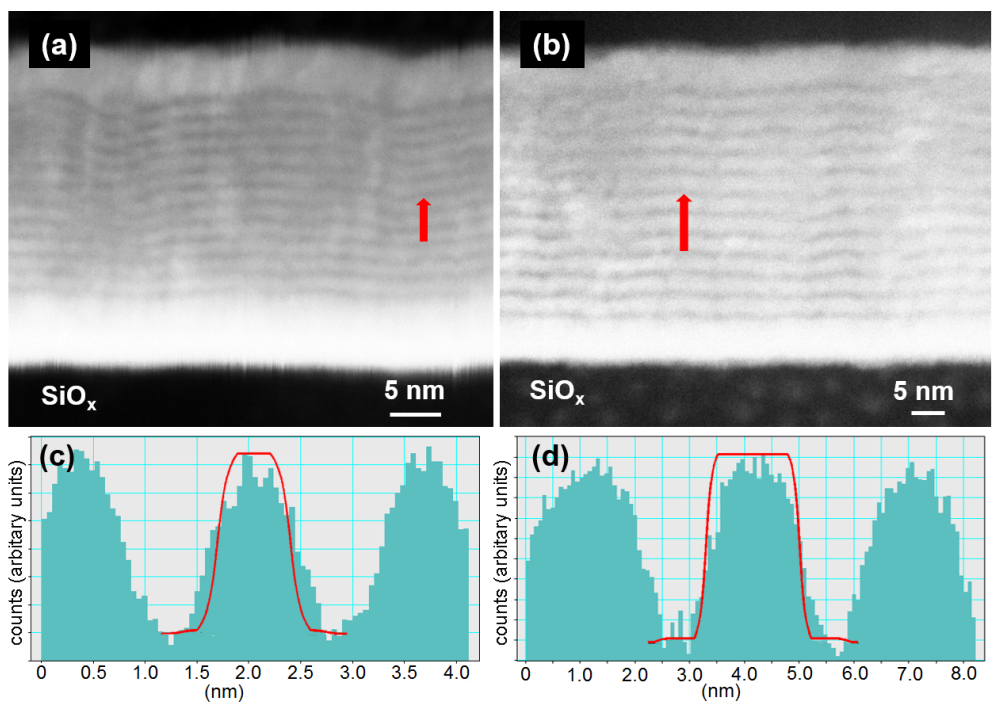

**Fig. 3.** Cross-section HAADF images of CoFe/Pd multilayer samples with (a) $t_{CoFe}$ = 0.45 nm, and (b) $t_{CoFe}$ = 1 nm. (c) & (d) Line profiles of the intensity positions and directions as indicated by red arrows in (a) & (b). Simulated intensity profiles are inserted as bold lines.

## 3.3 Magnetic domain structure

Off-axis electron holography is an interferometric electron–microscope technique that provides quantitative access to phase shifts experienced by the incident electron wavefront due to interactions with the electrostatic and magnetic potentials of the sample [16]. The changes in phase of the electron wave deduced from an electron hologram can then be used to provide quantitative information about the magnetic fields within and outside the sample with a spatial resolution that can approach the nanometer scale under optimal conditions [17-18]. Figure 4(a) shows a typical example of a reconstructed phase map from the sample with $t_{CoFe} = 0.45$ nm, where the magnitude of the local phase change is represented by changes in color. No clear information is available from within the multilayer region (labelled ML) because of phase-wrapping issues [18]. However, two magnetic domains are clearly visible in the amplified (×8) phase image in Fig. 4(b), showing magnetization perpendicular to the sample surface. Line profiles of the phase were made in the vacuum region near the top surfaces of the multilayers, and on the silicon oxide film near the bottom of the multilayers, as indicated by the dotted lines in Fig. 4(a). These phase profiles are shown in Figs. 4(c) and (d). The positions of magnetic domain walls can be identified from these line profiles, as indicated by the arrows. The magnitude of the magnetic field in projection related to the magnetic domains can also be estimated. Moreover, it can be seen that the magnetic fields emerging from the top and bottom surfaces of the multilayers are basically identical. Further observations of the samples with larger $t_{CoFe}$ revealed similar magnetic domain structure, except that these multilayers generated weaker magnetic field perpendicular to the sample surface. Perpendicular magnetic field was revealed by holography for the sample with $t_{CoFe} = 1$ nm but not apparent in the MFM images. In addition to the enhanced detection limit of holography compared with MFM [19], it is also possible that additional shape anisotropy resulting from cross-sectioning of the continuous film caused partial reorientation of the in-plane magnetization of the continuous film to the perpendicular direction.

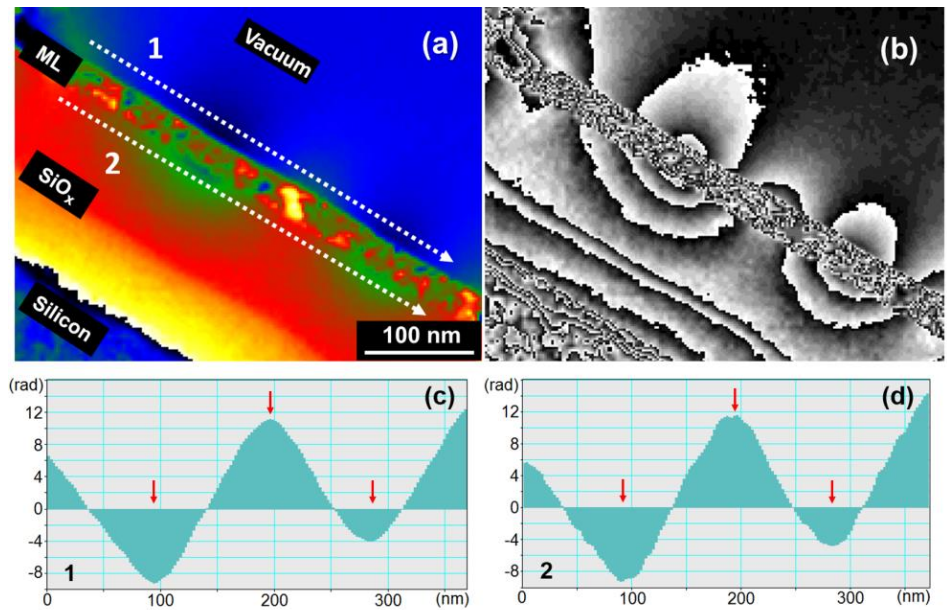

**Fig. 4.** (a) Reconstructed holography phase map of multilayers with $t_{CoFe} = 0.45$ nm, (b) Amplified (×8) phase image showing distribution of magnetic domains. (c) and (d) Line profiles from positions 1 and 2, respectively, marked in (a) by white arrows.

These holography experiments were done on cross-sectional TEM samples, so it might be anticipated that the revealed magnetic domain structure could be affected by the lamellar-like sample geometry. Indeed, our observation revealed that the apparent sizes of the magnetic domains changed slightly with the projected sample thickness along the electron-beam direction. The magnetic domain structures observed in phase maps, such as Fig. 4, are different from the structures observed by MFM, as shown in Fig. 1. Thus, the TEM sample geometry has some effect on the apparent magnetic domain size because the domains are constrained by the shorter plane of the multilayers rather than being randomly distributed. Figure 5 shows reconstructed phase maps for the sample with $t_{CoFe}$ = 0.8 nm indicating slight changes in the magnetic domain structure from areas with different projected sample thicknesses. The thickness values were estimated using the reconstructed amplitude image and the known inelastic-mean-free-path of silicon (±10% error estimate) [20-21]. The magnetic domains have average widths of 116 nm, 120 nm and 130 nm for areas with projected thicknesses of 412 nm, 525 nm and 729 nm.

Figure 6(a) shows another example of a phase map in the vacuum near the top surface of the sample with $t_{CoFe}$ = 0.45 nm. Contour lines corresponding to two magnetic domains are visible in the phase-amplified map shown in Fig. 6(b). These contour lines indicate differences in magnetic domain size, which could be caused either by different values of the magnetic field in projection or by rotation of the magnetic domains. In comparison, for the sample with $t_{CoFe}$ = 0.8 nm, as shown in Fig. 6(c) and (d), contour lines generated from different magnetic domains showed almost the same morphology, indicating less variability. Considering that PMA has an orientation that follows the multilayer stacking, layer waviness in the sample with thinner $t_{CoFe}$ is likely to cause variation among magnetic domains. Studies carried out by ferromagnetic resonance (FMR) also indicate that magnetic inhomogeneity increases as the anisotropy increases [22]. This result agrees with our observations because the PMA for these multilayers increases as $t_{CoFe}$ decreases

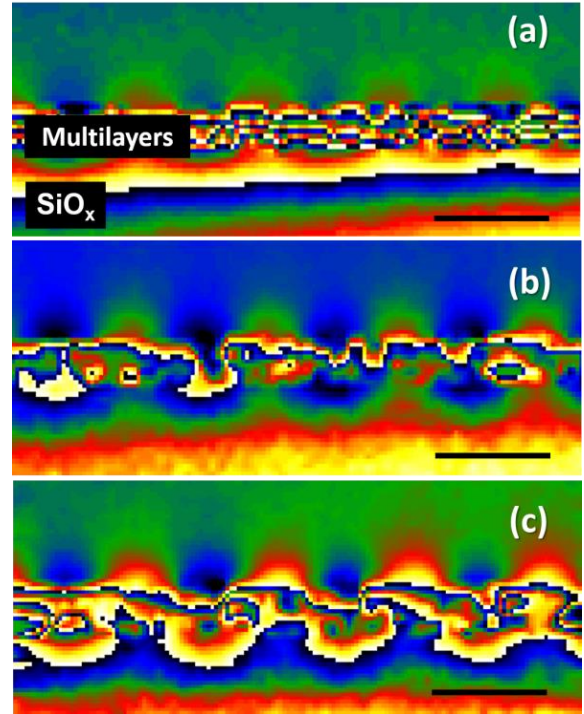

**Fig. 5.** Phase maps for CoFe/Pd multilayer sample with $t_{CoFe}$ = 0.8 nm. Sample thicknesses in beam direction are: (a) 412 nm, (b) 525 nm, (c) 729 nm. Scale bar = 100 nm in each case.

To further investigate the relationship between magnetic domain structure and microstructure, the positions of magnetic domain walls were marked on Lorentz images taken from the same areas where phase maps were generated. Although Lorentz images have limited resolution and less intensity than conventional TEM images, columns with different crystalline orientations are still visible. Figure 7 shows a Lorentz image from the sample with $t_{CoFe}$ = 1 nm, with the positions of magnetic domain walls indicated. The phase map used to locate these domains is also shown as an insert. As illustrated in Fig. 7, two magnetic domains with almost equal size are visible. In this case, there is no apparent correlation between the positions of magnetic domain walls and columns, although the crystal structure within the two areas covered by these two magnetic domains seems different.

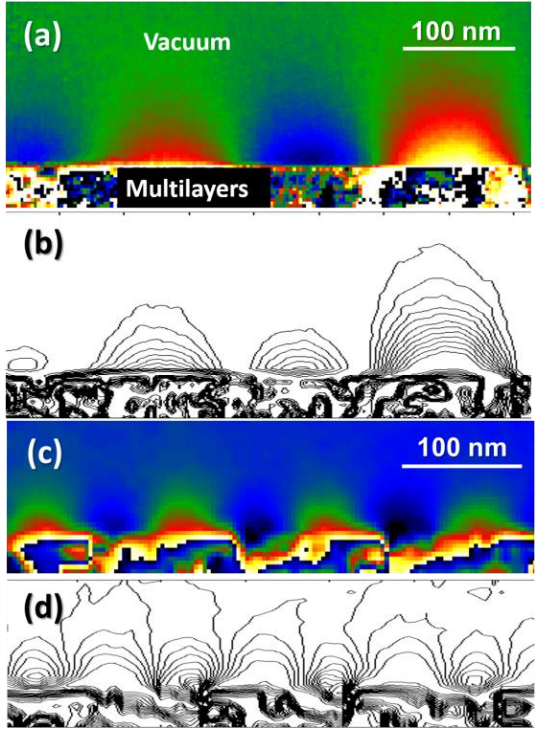

**Fig. 6.** (a) Phase map for CoFe/Pd multilayers sample with $t_{CoFe}$ = 0.45 nm. (b) Modified (cos) phase image of (a) showing contour lines; (c) Phase map of multilayers with $t_{CoFe}$ = 0.8 nm. (b) Modified (cos) phase image of (c) showing contour lines.

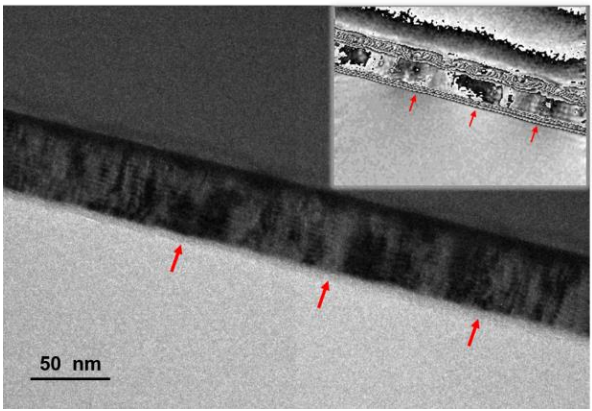

**Fig. 7.** Lorentz image for CoFe/Pd multilayers sample with $t_{CoFe}$ = 1 nm, magnetic domain walls in imaging area are indicated by red arrows. Phase map used for locating domain wall shown as insert.

## 4. Conclusions

Electron holography in conjunction with other characterization techniques has been used to investigate the variation of domain structure and perpendicular magnetic anisotropy in CoFe/Pd multilayers of different magnetic layer thickness ($t_{CoFe}$) in the range of 0.45 nm to 1.0 nm. All films showed a textured columnar morphology with some chemical interdiffusion between layers. In addition, for smaller $t_{CoFe}$, it was observed: i) The average columnar width within the film decreased but not impact magnetic domain formation; ii) Layer waviness and surface roughness increased; iii) More relative interdiffusion between multilayers; iv) The magnetic domain size and the magnetic field originating from each domain perpendicular to the sample surface increased; v) Variability between the magnetic domains increased. Thus these results confirm that bilayer thickness is the predominant factor affecting the magnetic domain structure and PMA. It is known that factors such as surface roughness, structure inhomogeneity and interdiffusion caused by decreasing $t_{CoFe}$ could deteriorate PMA [23-27], but this study confirmed that the sample with the thinnest CoFe layer of 0.45 nm still showed the strongest PMA. Overall, these results indicate the usefulness of off-axis electron holography as an alternative approach for characterizing the domain structure and PMA of magnetic multilayers, although the possibility of changes caused by sample cross-sectioning will still need to be taken into account.

**Acknowledgments**

This work was supported by DoE Grant DE-FG02-04ER46168. We gratefully acknowledge the use of facilities within the John M. Cowley Center for High Resolution Electron Microscopy at Arizona State University, and we thank the assistance from the M02 Group, Institute of Physics, Chinese Academy of Science for use of the AFM/MFM facility and discussion.